\documentclass[preprint,12pt,authoryear]{elsarticle}

\usepackage{amssymb}
\usepackage{latexsym}
\usepackage{amsmath}
\usepackage{color}
\usepackage{graphicx,dblfloatfix}
\usepackage{subfigure}
\usepackage[round]{natbib}
\usepackage[margin=1in]{geometry}
\usepackage{booktabs}
\usepackage{lscape}

\journal{Icarus}

\begin{document}

\begin{frontmatter}


%% \cortext[cor1]{}a

\title{An analytical model of crater count equilibrium}


\author{Masatoshi Hirabayashi$^1$, David A. Minton$^1$, Caleb I. Fassett$^2$}

\address{$^1$Earth, Atmospheric and Planetary Sciences, Purdue University, 550 Stadium Mall Drive,  Purdue University, West Lafayette, IN 47907-2051 United States}
\address{$^2$NASA Marshall Space Flight Center, Huntsville, AL 35805 United States}

\begin{abstract}
Crater count equilibrium occurs when new craters form at the same rate that old craters are erased, such that the total number of observable impacts remains constant. Despite substantial efforts to understand this process, there remain many unsolved problems. Here, we propose an analytical model that describes how a heavily cratered surface reaches a state of crater count equilibrium. The proposed model formulates three physical processes contributing to crater count equilibrium: cookie-cutting (simple, geometric overlap), ejecta-blanketing, and sandblasting (diffusive erosion). These three processes are modeled using a degradation parameter that describes the efficiency for a new crater to erase old craters. The flexibility of our newly developed model allows us to represent the processes that underlie crater count equilibrium problems. The results show that when the slope of the production function is steeper than that of the equilibrium state, the power law of the equilibrium slope is independent of that of the production function slope. We apply our model to the cratering conditions in the Sinus Medii region and at the Apollo 15 landing site on the Moon and demonstrate that a consistent degradation parameterization can successfully be determined based on the empirical results of these regions. Further developments of this model will enable us to better understand the surface evolution of airless bodies due to impact bombardment. 
\end{abstract}

\begin{keyword}
Cratering \sep Impact processes \sep Regoliths


\end{keyword}

\end{frontmatter}


\section{Introduction}
A surface's crater population is said to be in equilibrium when the terrain loses visible craters at the same rate that craters are newly generated \citep[e.g.,][]{Melosh1989, Melosh2011}. Since the Apollo era, a number of studies have been widely conducted that have provided us with useful empirical information about crater count equilibrium on planetary surfaces. In a cumulative size-frequency distribution (CSFD), for many cases, the slope of the equilibrium state in log-log space ranges from -1.8 to -2.0 if the slope of the crater production function is steeper than -2 \citep{Gault1970, Hartmann1984, Chapman1986, Xiao2015}. 

\begin{figure}[!]
  \centering
  \includegraphics[width=\textwidth]{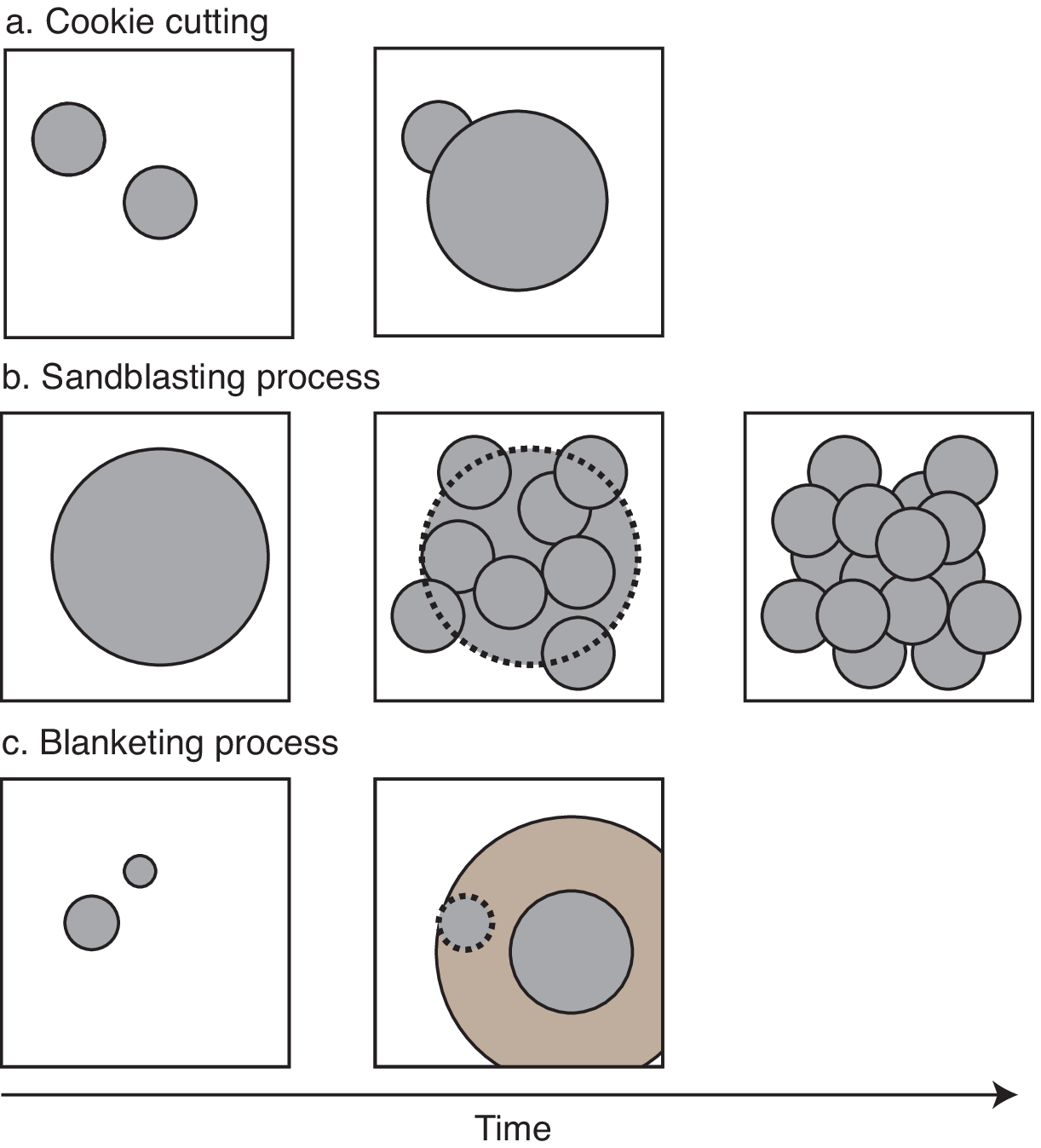}
  \caption{A schematic plot of the processes that make craters invisible. \textbf{a}, The cookie-cutting process, where each new crater overprints older craters. \textbf{b}, The sandblasting process, where multiple small craters collectively erode a larger crater. \textbf{c}, The blanketing process, where ejecta from a new crater buries old craters. The brown circle with the solid line shows ejecta blankets. The gray circles with the solid lines indicate fresh craters, and the gray circles with the dashed lines describe partially degraded craters.}
  \label{Fig:Equilibrium4}
\end{figure}

When new craters form, they erase old craters. There are three crater erasure processes that primarily contribute to crater count equilibrium (Figure \ref{Fig:Equilibrium4}). The first process is cookie-cutting, in which a new crater simply overlaps old craters. Complete overlap can erase the craters beneath the new crater. However, if overlapping is incomplete, the old craters may be still visible because the rim size of the new crater strictly restricts the range of cookie-cutting. Cookie-cutting is a geometric process and only depends on the area occupied by new craters. 

However, since craters are not two-dimensional circles but three-dimensional depressions, cookie-cutting is ineffective when a newly-formed crater is smaller than a pre-existing crater beneath it. The second process considers this three-dimensional effect. This process is sometimes called sandblasting\footnote{Earlier works called this process small impact erosion \citep{Ross1968, Soderblom1970}. Here, we follow the terminology by \cite{Minton2015}.}, which happens when small craters collectively erode a large crater by inducing downslope diffusion, thus filling in the larger depression over time \citep{Ross1968, Soderblom1970, Fassett2014}. 

Lastly, blanketing by ejecta deposits covers old craters outside the new crater rim \citep[e.g.,][]{Fassett2011}. The thickness of ejecta blankets determines how this process contributes to crater count equilibrium. However, it has long been noted that they are relatively inefficient at erasing old craters \citep{Woronow1977}. For these reasons, this study formulates the ejecta-blanketing process as a geometric overlapping process (like cookie-cutting) and neglects it in the demonstration exercise of our model. 

To our knowledge, \cite{Gault1970} is the only researcher known to have conducted comprehensive laboratory-scale demonstrations for the crater count equilibrium problem. In a 2.5-m square box filled 30-cm deep with quartz sand, he created six sizes of craters to generate crater count equilibrium. The photographs taken during the experiments captured the nature of crater count equilibrium (Figure 5 in \cite{Gault1970}). His experiments successfully recovered the equilibrium level of crater counts observed in heavily cratered lunar terrains. 

Earlier works conducted analytical modeling of crater count equilibrium \citep{Marcus1964, Marcus1966, Marcus1970, Ross1968, Soderblom1970, Gault1974mixing}. \cite{Marcus1964, Marcus1966, Marcus1970} theoretically explored the crater count equilibrium mechanism by considering simple circle emplacements. The role of sandblasting in crater erosion was proposed by \cite{Ross1968}, followed by a study of sandblasting as an analog of the diffusion problem \citep{Soderblom1970}. Each of these models addressed the fact that the equilibrium slope was found to be 2, which does not fully capture the observed slope described above. \cite{Gault1974mixing} also developed a time-evolution model for geometric saturation for single sized craters. 

Advances in computers have made Monte-Carlo simulation techniques popular for investigating the evolution of crater count equilibrium. Earlier works showed how such techniques could describe the evolution of a cratered surface \citep{Woronow1978}. Since then, the techniques have become more sophisticated and have been capable of describing complicated cratering processes \citep{Hartmann1997, Marchi2014, Richardson2009, Minton2015}. Many of these Monte Carlo codes represent craters on a surface as simple circles or points on a rim \citep{Woronow1977, Woronow1978, Chapman1986, Marchi2014}. The main drawback of the earlier analytical models and some Monte-Carlo techniques that simply emplaced circles was that crater erasure processes did not account for the three-dimensional nature of crater formation. As cratering proceeds, diffusive erosion becomes important \citep{Fassett2014}. Techniques that model craters as three-dimensional topographic features capture this process naturally \citep{Hartmann1997,Minton2015}; however, these techniques are computationally expensive.  

Here, we develop a model that accounts for the overlapping process (cookie-cutting and ejecta-blanketing) and  the diffusion process (sandblasting) to describe the evolution of crater count equilibrium while avoiding the drawbacks discussed above. Similar to \cite{Marcus1964, Marcus1966, Marcus1970}, we approach this problem analytically. In the proposed model, we overcome the computational uncertainties and difficulties that his model encountered, such as his complex geometrical formulations. Since the proposed model has an analytical solution, it is efficient and can be used to investigate a larger parameter space than numerical techniques. Although a recent study reported a model of the topographic distribution of cratered terrains, which have also reached crater count equilibrium at small crater sizes, on the Moon \citep{Rosenburg2015}, the present paper only focuses on the population distribution of craters. We emphasize that the presented model is a powerful tool for considering the crater count equilibrium problem for any airless planets. In the present exercise, we consider a special size range between 0 and $\infty$ to better understand crater count equilibrium. Also, although the crater count equilibrium slope may potentially depend on the crater radius, we assume that it is constant.

We organize the present paper as follows. In Section \ref{Sec:SFD}, we introduce a mathematical form that describes the produced crater CSFD derived from the crater production function. Section \ref{Sec:Modelling} provides the general formulation of the proposed model. This form will be related to the form derived by \cite{Marcus1964} but will be more flexible to consider the detailed erasure processes. In Section \ref{Sec:Generalformulation}, we derive an analytical solution to the derived equation. In Section \ref{Sec:Behavior}, we apply this model to the crater count equilibrium problems of the Sinus Medii region and the Apollo 15 landing site on the Moon, and demonstrate how we can use observational crater counts to infer the nature of the crater erasure processes.

\section{The produced crater CSFD}
\label{Sec:SFD}
To characterize crater count equilibrium, we require two populations: the crater production function and the produced craters. The production function is an idealized model for the population of craters that is expected to form on a terrain per time and area. In this study, we use a CSFD to describe the produced craters. The crater production function in a form of CSFD is given as \citep{Crater1979standard}
\begin{eqnarray}
P_{(\ge r)} = \sigma \hat x r^{-\eta}, \label{Eq:Ctl}
\end{eqnarray}
where $\eta$ is the slope\footnote{For notational simplification, we define the slopes as positive values.}, $\sigma$ is a constant parameter with units of m$^{\eta-2}$ s$^{-1}$, $\hat x$ is a cratering chronology function, which is defined to be dimensionless (e.g., for the lunar case, \cite{Neukum2001}), and $r$ is the crater radius. For notational simplification, we choose to use the radius instead of the diameter. 

In contrast, the produced craters are those that actually formed on the surface over finite time in a finite area. If all the produced craters are counted, the mean of the produced crater CSFD over many samplings is obtained by factorizing the crater production function by a given area, $A$, and by integrating it over time, $t$. We write the produced crater CSFD as
\begin{eqnarray}
C_{t (\ge r)} =  A \sigma r^{-\eta} \int_0^{t} \hat x dt = A \xi r^{-\eta} \int_0^t x dt =  A \xi X r^{-\eta},\label{Eq:C_t}
\end{eqnarray}
where
\begin{eqnarray}
\xi &=& \sigma \int_{0}^{t_s} \hat x dt \:\: \text{[m$^{\eta-2}$}], \\ 
x &=& \frac{\hat x}{\int_{0}^{t_s} \hat x dt} \:\: \text{[s$^{-1}$}], \label{Eq:x} \\
X &=& \int_0^t x dt. \label{Eq:CapX}
\end{eqnarray} 
Consider a special case that the cratering chronology function is constant, for example, $\hat x = 1$. For this case, $x  =  1/t_s$ and $X = t / t_s$. In these forms, $t_s$ can be chosen arbitrarily. For instance, it is convenient to select $t_s$ such that $A \xi r^{-\eta}$ recovers the produced crater CSFD of the empirical data at $X = 1$. Also, we set the initial time as zero without losing the generality of this problem. Later, we use $x$ and $X$ in the formulation process below.

Given the produced crater CSFD, $C_{t (\ge r)}$, the present paper considers how the visible crater CSFD, $C_{c (\ge r)}$, evolves over time. The following sections shall omit the subscript, $(\ge r)$, from the visible crater CSFD and the produced crater CSFD to simplify the notational expressions. 

\section{Development of an analytical model}
\label{Sec:Modelling}

\subsection{The concept of crater count equilibrium}
\label{Sec:Concept}
We first introduce how the observed number of craters on a terrain subject to impact bombardment evolves over time. The most direct way to investigate this evolution would be to count each newly generated and erased craters over some interval of time. This is what Monte Carlo codes, like CTEM, do \citep{Richardson2009, Minton2015}. In the proposed model, the degradation processes are parameterized by a quantity that describes how many craters are erased by a new crater. We call this quantity the degradation parameter. 

Consider the number of visible craters of size $i$ on time step $s$ to be $N_i^s$. We assume that the production rate of craters of this size is such that exactly one crater of this size is produced in each time step. In our model, we treat partial degradation of the craters by introducing a fractional number. Note that since we do not account for the topological features of the cratered surface in the present version, this parameter does not distinguish degradation effects on shapes such as a rim fraction and change in a crater depth. This consideration is beyond our scope here. For instance, if $N_i^2=1.5$, this means that at time $2$, the first crater is halfway to being uncounted.\footnote{Because $N^0_i = 0$, at time 1 there is only one crater emplaced.} If there is no loss of craters, $N_i^s$ should be equal to $s$. When old craters are degraded by new craters, $N_i^s$ becomes smaller than $s$. We describe this process as
\begin{eqnarray}
N_i^s = N_{i}^{s-1} + 1 - \Omega_i k_i^s N_i^{s-1}, \label{Eq:Nij1}
\end{eqnarray}
where $k_i^s$ is the degradation parameter describing how many craters of size $i$ lose their identities (either partial or in full) on time step $s$, and $\Omega_i$ is a constant factor that includes the produced crater CSFD and the geometrical limits for the $i$th-sized craters. In the analytical model, we average $k^s_i$ over the total number of the produced craters. This operation is defined as
\begin{eqnarray}
k_i = \frac{\sum_{s = 1}^{s_{max}} k_i^s N_i^{s-1}}{\sum_{s = 1}^{s_{max}} N_i^{s-1}}. \label{Eq:bar_kappa}
\end{eqnarray}
Equation (\ref{Eq:bar_kappa}) provides the following form:
\begin{eqnarray}
N_i^s = N_{i}^{s-1} + 1 - \Omega_i k_i N_i^{s-1}. \label{Eq:Nij2}
\end{eqnarray}
In the following discussion, we use this averaged value, $k_i$, and simply call it the degradation parameter without confusion. 

\subsection{The case of a single sized crater production function.}
\label{Sec:singleSize}
To help develop our model, we first consider a simplified case of a terrain that is bombarded only by craters with a single radius. This case is similar to the analysis by \cite{Gault1974mixing}. We consider the number of craters, instead of the fraction of the cratered area that was used by \cite{Gault1974mixing}.  

Consider a square area in which single sized craters of radius, $r_i$, are generated and erased over time. $n_i$ is the number of the produced craters, $A$ is the area of the domain, $N_i $ is the number of visible craters at a given time, and $N_{0,i}$ is the maximum number of visible craters of size $i$ that is possibly visible on the surface (geometric saturation). In the following discussion, we define the time-derivative of $n_i$ as $\dot n_i$. $N_i $ is the quantity that we will solve, and $N_{0,i}$ is defined as
\begin{eqnarray}
N_{0,i} = \frac{A q}{\pi r_i^2}, \label{Eq:Nt_def}
\end{eqnarray}
where $q$ is the geometric saturation factor, which describes the highest crater density that could theoretically be possible if the craters were efficiently emplaced onto the surface in a hexagonal configuration \citep{Gault1970}. For single sized circles $q=\pi / 2 \sqrt{3} \sim 0.907$. 

The number of visible craters will increase linearly with time at rate, $\dot n_i $, at the beginning of impact cratering. After a certain time, the number of visible craters will be obliterated with a rate of $k_i \dot n_i N_i/N_{0,i}$, where $N_i/N_{0,i}$ means the probability that a newly generated crater can overlap old craters. For this case, $\Omega_i = \dot n_i/N_{0,i}$, and $k_i$ represents the number of craters erased by one new crater at the geometrical saturation condition. This process provides the first-order ordinal differential equation, which is given as
\begin{eqnarray}
\frac{d N_i }{d t}  = \dot n_i  - k_i \dot n_i  \frac{N_i }{N_{0,i}}. \label{Eq:eqmSingle}
\end{eqnarray}
The initial condition is $N_i = 0$ at $ t = 0$. The solution of Equation (\ref{Eq:eqmSingle}) is given as 
\begin{eqnarray}
N_i &=& \frac{N_{0,i}}{k_i} \left \{1-  \exp \left( - \frac{k_i n_i}{N_{0,i}} \right) \right \}, \nonumber \\
&=& \frac{A q}{k_i \pi r_i^2} \left[ 1-  \exp \left \{ - \frac{k_i \pi r_i^2 n_{i}}{A q} \right \} \right]. \label{Eq:Nsol}
\end{eqnarray}

\begin{figure}[!]
  \centering
  \includegraphics[width=\textwidth]{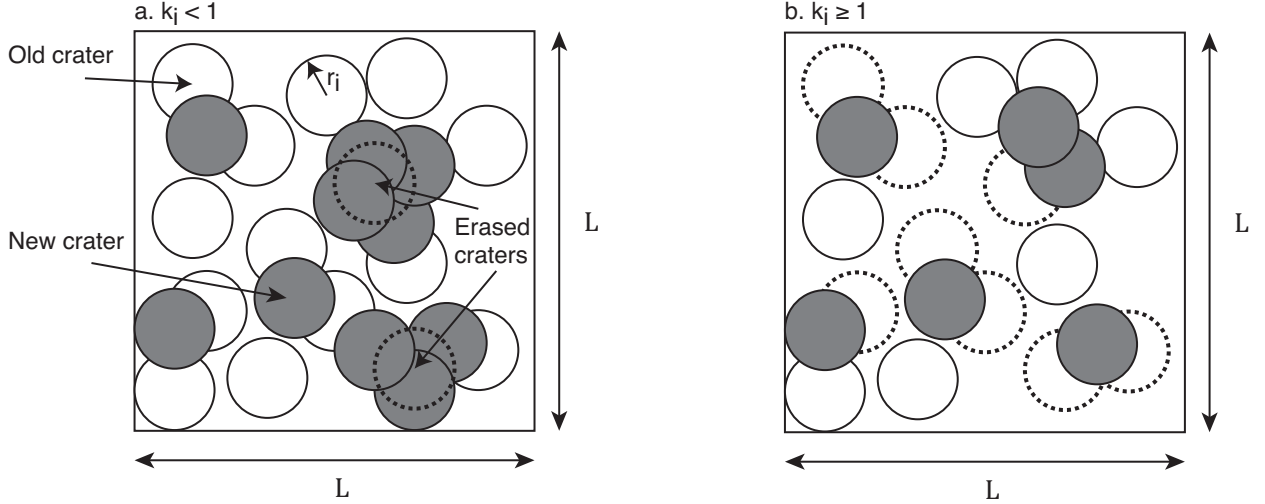}
  \caption{The physical meaning of the degradation parameter for the case of a single sized crater production function. The gray circles are newly emplaced craters, while the white circles are old craters. The solid circles represent visible craters, and the dashed circles represent lost craters. \textbf{a}, The case where 2 old craters were completely lost and several ones were partially erased after 10 new craters formed ($k_i < 1$). \textbf{b}, The case where 8 old craters were completely lost and several ones were partially erased after 6 new craters formed ($k_i \ge 1$).}
  \label{Fig:Single}
\end{figure}

To proceed further, we must understand the physical meaning of the degradation parameter. Equation (\ref{Eq:eqmSingle}) indicates that the degradation parameter represents how many old craters are erased by a new crater. Figure \ref{Fig:Single} shows two examples that describe different degradation parameters in the case of a single crater-size production function. New craters are represented as gray circles, old visible craters are white circles with solid borders, and lost old craters are white circles with dashed borders. If $k_i$ is less than 1, more new craters are necessary to be emplaced to erase old craters. If $k_i$ is larger than 1, one new crater can erase more than one old crater. From Equation (\ref{Eq:Nsol}), as $t \rightarrow \infty$, $N_i$ reaches $N_{0,i}/k_i$, not $N_{0,i}$. Thus, since $N_{0,i} / k_i \le N_{0,i}$, $k_i \ge 1$. This means that the case described in Figure \ref{Fig:Single}a, $k_i < 1$, does not happen. 

\subsection{The case of a multiple crater size production function.} 
\label{Sec:MultipleSize}
\begin{figure}[!]
  \centering
  \includegraphics[width=\textwidth]{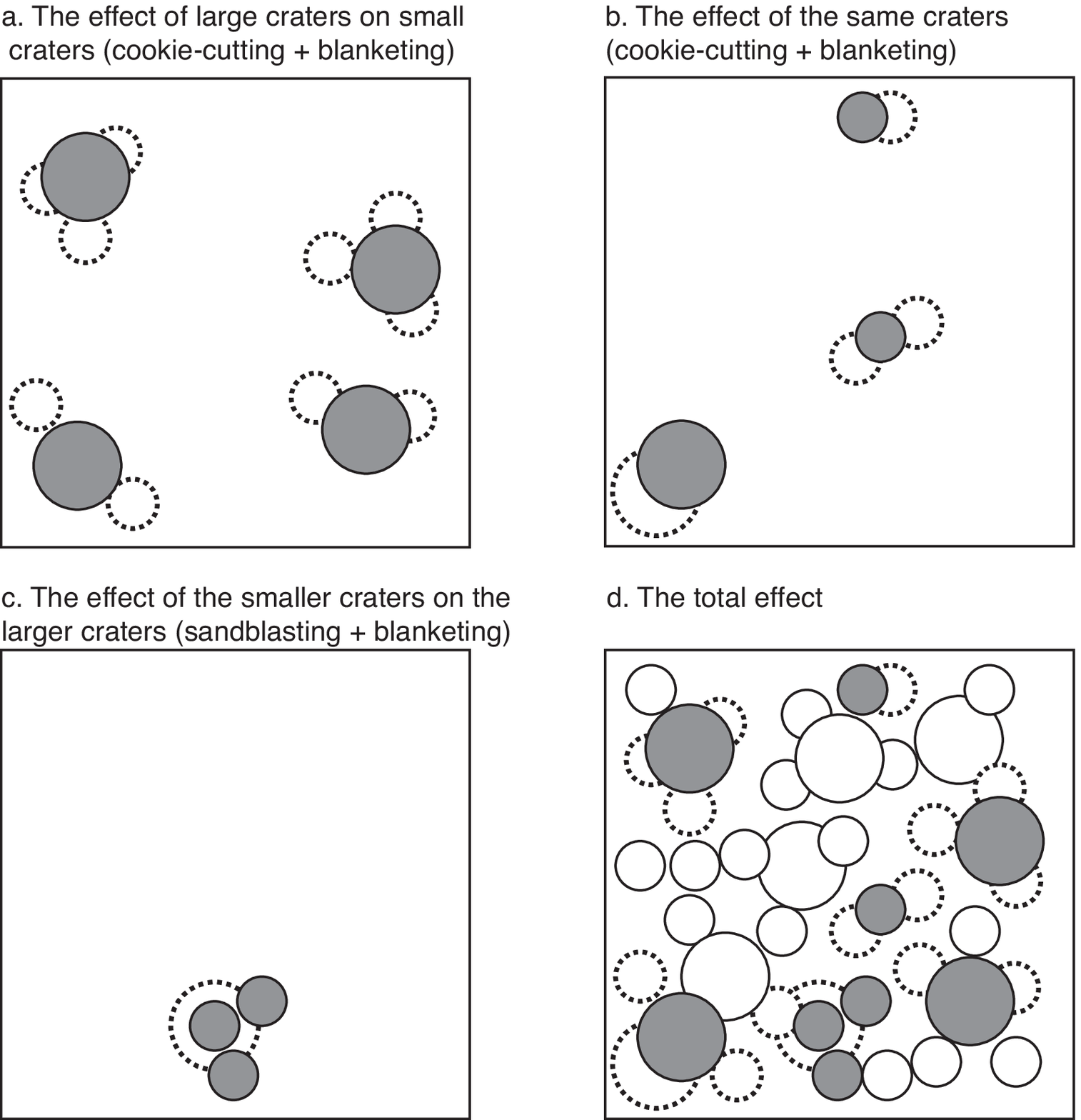}
  \caption{Schematic plot of degradation processes for a crater production function with two crater sizes, large and small. The model accounts for three different cases to describe the total effect. \textbf{a}, The effect of large craters on small craters. \textbf{b}, The effect of the same-sized craters. \textbf{c}, The effect of small craters on large craters. \textbf{d}, The total effect obtained by summing the effects given in \textbf{a} through \textbf{c}. The gray circles are new craters, the white circles with solid lines describe visible old craters, and the white circles with dashed lines indicate invisible old craters.}
  \label{Fig:Multi}
\end{figure}

\begin{figure}[!]
  \centering
  \includegraphics[width=\textwidth]{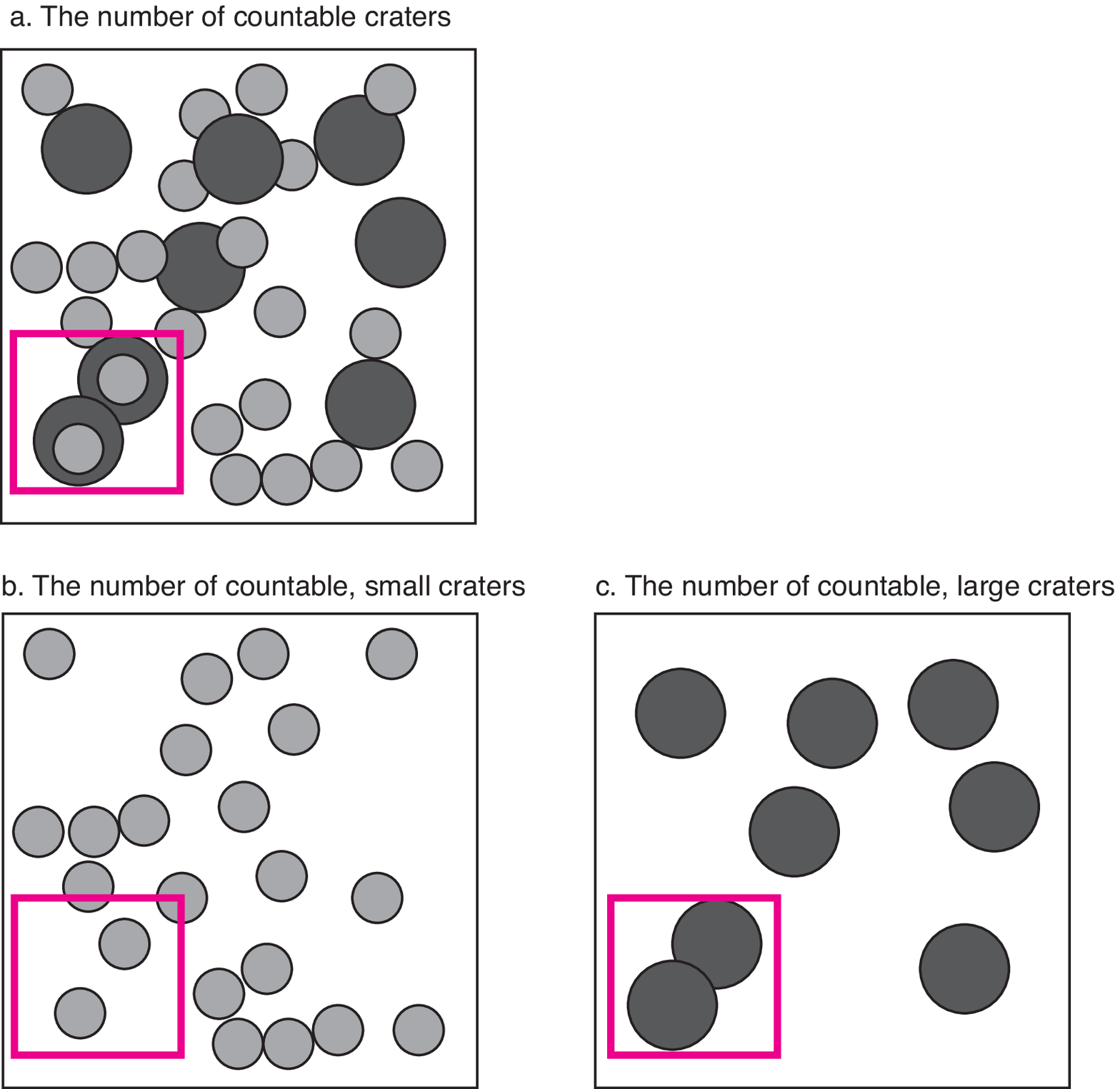}
  \caption{Schematic plot for how the analytical model computes the number of visible craters for the multiple crater-size case. The model tracks the number of visible craters for each size and sums up that of all the considered sizes. In case small craters are emplaced on larger crates (e.g., the solid square), the model counts both sizes and sums it up to compute the CSFD. \textbf{a}, The number of visible craters that the model is supposed to count. \textbf{b} and \textbf{c}, The crater counting for each case. The light and dark gray circles show small and large craters, respectively.}
  \label{Fig:Overlapping}
\end{figure}

This section extends the single crater-size case to the multiple crater-size case. In the analytical model by \cite{Marcus1964, Marcus1966, Marcus1970}, complex geometric considerations were necessary, and there were many uncertainties. We will see that the extension of Section \ref{Sec:singleSize} makes our analytical formulation clear and flexible so that the developed model can take into account realistic erasure processes. We create a differential equation for this case based on Equation (\ref{Eq:eqmSingle}), which states that for a given radius, the change rate of visible craters is equal to the difference between the number of newly emplaced craters per time and that of newly erased craters per time.

We first develop a discretized model and then convert it into a continuous model. In the discussion, we keep using the notations defined in Section \ref{Sec:singleSize}. That is, for each $i$th crater, the radius is $r_i$, the number of visible craters is $N_i$, the maximum number of visible craters in geometric saturation is $N_{0,i}$, the crater production rate is $\dot n_i $, and the total number of the produced craters is $n_i$. Craters of size $i$ are now affected by those of size $j$, and we define these quantities for craters of size $j$ in the same way. 

Modeling the degradation process of differently sized craters starts from formulating how the number of visible craters of one size changes due to craters of other sizes on each step. Figure \ref{Fig:Multi} shows how the number of visible craters changes based on the degradation processes that operate during the formation of new craters. In this figure, we illustrate the cratering relation between small craters and large craters to visualize the process clearly. For this case, there are three possibilities. First, new large craters erase smaller, older ones by either cookie-cutting or ejecta-blanketing (Figure \ref{Fig:Multi}a). Second, new craters eliminate the same-sized craters by either cookie-cutting or ejecta-blanketing (Figure \ref{Fig:Multi}b). Finally, new small craters can degrade larger ones through sandblasting or ejecta-blanketing (Figure \ref{Fig:Multi}c). Figure \ref{Fig:Multi}d shows the total effect when all possible permutations are considered.

Consider the change in the number of visible craters of size $i$. The $i$th-sized craters are generated with the rate, $n_i $, on every time step. This accumulation rate is provided as
\begin{eqnarray}
\left.\frac{d N_i }{d t}\right|_{acc} &=& \dot n_i. \label{Eq:ith_interact}
\end{eqnarray}
The present case requires consideration of differently sized craters. The degradation parameter should vary according to the degradation processes of the $j$th-sized craters. To account for them (Figure \ref{Fig:Multi}), we define the degradation parameter describing the effect of the $j$th-sized craters on the $i$th-sized craters as $k_{ij}$. Given the $i$th-sized craters, we give the degradation rate due to craters of size $j$ as 
\begin{eqnarray}
\left.\frac{d N_i }{d t}\right|_{deg, j} &=& - k_{ij} \dot n_j  \frac{N_i }{N_{0,i}} \frac{r_j^{2}}{r_i^2}. \label{Eq:jth_interact}
\end{eqnarray}
In this equation, $\Omega_i$ in Equation (\ref{Eq:Nij1}) becomes dependent on craters of size $j$. By defining this function for this case as $\Omega_{ij}$, we write 
\begin{eqnarray}
\Omega_{ij} = \frac{\dot n_j }{N_{0,i}} \frac{r_j^2}{r_i^2}. 
\end{eqnarray}
$k_{ij}$ is a factor describing how effectively craters of size $i$ are erased by one crater of size $j$. For example, when $k_{ij} = 1$, $r_j^2 / r_i^2$ is the total number of craters of size $i$ erased by one crater of size $j$ at the geometric saturation condition. In the following discussion, we constrain $k_{ij}$ to be continuous over the range that includes $r_i = r_j$ to account for the degradation relationships among any different sizes.

Based on Equation (\ref{Eq:ith_interact}) and (\ref{Eq:jth_interact}), we take into account both the accumulation process and the degradation process. Considering the possible range of the crater size, we obtain the first-order differential equation for the time evolution of $N_i$ as
\begin{eqnarray}
\frac{d N_i }{d t} &=& \left.\frac{d N_i }{d t}\right|_{acc} + \sum_{j = i_{min}}^{i_{max}} \left.\frac{d N_i }{d t}\right|_{deg,j}, \nonumber \\
&=& \dot n_{i}  - \frac{N_i }{N_{0,i}}  \sum_{j = i_{min}}^{i_{max}} k_{ij} \dot n_j \frac{r_j^2}{r_i^2}. \label{Eq:N_i}
\end{eqnarray}
The summation operation on the right hand side means a sum from the largest craters to the smallest craters. Similar to the single-sized case, we set the initial condition such that $N_i = 0$ at $t = 0$. The solution of this equation is written as
\begin{eqnarray}
N_i = \frac{\dot n_i}{ \frac{\pi}{A q} \sum_{j = i_{min}}^{i_{max}} k_{ij} r_j^{2} \dot n_j} \left[ 1 - \exp \left( - \frac{\pi}{A q} \sum_{j = i_{min}}^{i_{max}} k_{ij} r_j^{2} n_j \right) \right]. \label{Eq:discrete}
\end{eqnarray}

As discussed in Section \ref{Sec:SFD}, it is common to use the CSFDs to describe the number of visible craters. To enable this model directly to compare its results with the empirical data, we convert Equation (\ref{Eq:discrete}) to a continuous form. $k_{ij}$ is rewritten as a continuous form, $k$. We write the continuous form of $r_i$ and that of $r_j$ as $r$ and $\check r$, respectively. We define $C_c$ as the CSFD of visible craters and rewrite $N_i$ and $n_i$ as
\begin{eqnarray}
N_i  \sim - \frac{d C_c }{d r} dr, \:\:\: n_i \sim - \frac{d C_t }{d r} dr,
\end{eqnarray}
respectively. Substitutions of these forms into Equation (\ref{Eq:discrete}) yields a differential form of the CSFD,
\begin{eqnarray}
\frac{d C_c}{d r} = - \frac{\frac{d \dot C_t}{d r}}{\frac{\pi}{Aq} \int_{r_{min}}^{r_{max}} \frac{d \dot C_t}{d \check r} k \check r^2 d \check r} \left[ 1 - \exp \left(\frac{\pi}{A q} \int_{r_{min}}^{r_{max}} \frac{d C_t}{d \check r} k \check r^2 d \check r \right) \right], \label{Eq:C_c}
\end{eqnarray}
where $r_{min}$ and $r_{max}$ are the smallest crater radius and the largest crater radius, respectively. The radius of the $i_{min}$th-sized craters and that of the $i_{max}$th-sized craters correspond to $r_{min}$ and $r_{max}$, respectively. Also, $\dot C_t$ is the time-derivative of $C_t$. In the following discussion, we will consider $r_{min} \rightarrow 0$ and $r_{max} \rightarrow \infty$ after we model the $k$ parameter. Equation (\ref{Eq:C_c}) is similar to Equation (18) in \cite{Marcus1964}, which is the key equation of his sequential studies. The crater birth rate, $\lambda$, and the crater damaging rate, $\mu$, are related to $- d \dot C_t / d r$ and $\frac{\pi}{Aq} \int_{r_{min}}^{r_{max}} \frac{d \dot C_t}{d \check r} k \check r^2 d \check r$, respectively. While his $\lambda$ and $\mu$ included a number of geometric uncertainties and did not consider the effect of three dimensional depressions on crater count equilibrium, our formulation overcomes the drawback of his model and provides much stronger constraints on the equilibrium state than his model. 

\section{Analytical solutions}
\label{Sec:Generalformulation}

\subsection{Formulation of the degradation parameter}
To determine a useful form of the degradation parameter, $k$, we start by discussing how this parameter varies as a function of $\check r$. If craters with a radius of $\check r$ are larger than those with a radius of $r$, cookie-cutting and ejecta-blanketing are main contributors to erasing the $r$-radius craters. In this study, we consider that cookie-cutting and ejecta-blanketing are only related to the geometrical relationship between craters with a radius of $\check r$ and those with a radius of $r$. Cookie-cutting only entails the geometric overlap of the $r$-radius craters; thus $k$ should always be one. Ejecta-blanketing makes additional craters invisible \citep{Pike1974, Fassett2011, Xie2016}, so $k$ is described by some small constant, $\alpha_{eb}$. Adding these values, we obtain the degradation parameter at $\check r \ge r$ as $1 + \alpha_{eb}$. 

If craters with a radius of $\check r$ are smaller than those with a radius of $r$, the possible processes that degrade the $r$-radius craters are ejecta-blanketing and sandblasting \citep{Fassett2014}. For simplicity, we only consider the size-dependence of sandblasting. It is reasonable that as $\check r$ becomes smaller, the timescale of degrading the $r$-radius crater should become longer \citep{Minton2015}. This means that with a small radius, the effect of new craters on the degradation process becomes small. To account for this fact, we assume that at $\check r < r$, $k$ increases as $\check r$ becomes large. Here, we model this feature by introducing a single slope function of $\check r/r$ whose power is a function of $r$.  

Combining these conditions, we define the degradation parameter as
\begin{eqnarray}
k =
  \begin{cases}
    (1 + \alpha_{eb}) \left( \frac{\check r}{r} \right)^{b (r)} & \quad \text{if } \check r < r, \\
    1 + \alpha_{eb}  & \quad \text{if } \check r \ge r, \\
  \end{cases} 
  \label{Eq:k}
\end{eqnarray}
where $b(r)$ is a positive function changing due to $r$. We multiplied $1 + \alpha_{eb}$ by the size-dependent term, $( \check r / r)^{b (r)}$, at $r < \check r$ for conveniency. This operation satisfies the continuity at $r = \check r$. 

We substitute the produced crater CSFD defined by Equation (\ref{Eq:C_t}) and the degradation parameter given by Equation (\ref{Eq:k}) into the integral term in Equation (\ref{Eq:C_c}). We rewrite the integral term of Equation (\ref{Eq:C_c}) as
\begin{eqnarray}
\int_{r_{min}}^{r_{max}} \frac{d C_t}{d \check r} k \check r^2 d \check r
&=& - \eta A \xi X (1 + \alpha_{eb}) \nonumber \\
&& \left \{\int_{r_{min}}^{r} \left( \frac{\check r}{r} \right)^{b (r)} \check r^{-\eta + 1} d\check r + \int_{r}^{r_{max}}  \check r^{-\eta + 1} d\check r \right \}, \nonumber \\
&=& - \eta A \xi X (1 + \alpha_{eb}) \nonumber \\
&& \left[ \frac{r^{ - \eta  + 2}}{-\eta + 2 + b (r)} \left \{ 1 - \left(\frac{r_{min}}{r} \right)^{-\eta + 2 + b(r)} \right \} \right. \nonumber \\
&& \left. +  \frac{r^{-\eta + 2}}{\eta - 2} \left \{ 1 - \left(\frac{r_{max}}{r} \right)^{-\eta + 2} \right \} \right]. \label{Eq:integral}
\end{eqnarray}
For the case of $\dot C_t$ that appears in the denominator of the fraction term in Equation (\ref{Eq:C_c}), we can use the derivation process above by replacing $X$ by $x$ (see Equation (\ref{Eq:CapX})). We have the similar operations below and only introduce the $C_t$ case without confusion. Note that the second operation in this equation is valid under the assumption that neither $-\eta + 2$ nor $-\eta + 2 + b(r)$ is zero. Under this condition, we examine whether or not Equation (\ref{Eq:integral}) has a reasonable value at $r_{min} \rightarrow 0$ and at $r_{max} \rightarrow \infty$. Later, we will show that an additional condition is necessary for $b(r)$ for $r_{min} \rightarrow 0$.

The term in the last row in Equation (\ref{Eq:integral}) has $(r_{max}/r)^{-\eta +2}$. If the slope of the produced crater CSFD satisfies $\eta - 2 > 0$, we obtain
\begin{eqnarray}
\left(\frac{r_{max}}{r} \right)^{-\eta + 2}  &<& 1. \label{Eq:cond1}
\end{eqnarray}
Thus, when $r_{max} \rightarrow \infty$, this term goes to zero. The term in the second to the last row in Equation (\ref{Eq:integral}) provides constraints on how the sandblasting process works to create the equilibrium states. Since $b(r) > 0$ and $\eta - 2 > 0$, the power of $r_{min}/r$, $-\eta + 2 + b(r)$, can only take one of the following cases: negative $(-\eta + 2 + b(r) < 0)$ or positive $(-\eta + 2 + b(r) > 0)$. If $-\eta + 2 + b(r) < 0$, the term, $(r_{min}/r)^{-\eta + 2 + b(r)}$, becomes $\infty$ at $r_{min} \rightarrow 0$. This condition yields $C_c \rightarrow 0$ at $r_{min} \rightarrow 0$. We rule out this condition by conducting the following thought experiment. We assume that this case is true. In nature, micrometeoroids play significant roles in crater degradation \citep{Melosh2011}. Because we assumed that this case is true, there should be no craters on the surface. This result obviously contradicts what we have seen on the surface of airless bodies (we see craters!).

The only possible case is the positive slope case, providing the condition that the sandblasting effect leads to crater count equilibrium as $b(r) > \eta - 2$. Since this case satisfies 
\begin{eqnarray}
\left(\frac{r_{min}}{r} \right)^{-\eta + 2 + b(r)} < 1, 
\end{eqnarray}
Equation (\ref{Eq:integral}) at $r_{min} \rightarrow 0$ and $r_{max} \rightarrow \infty$ is
\begin{eqnarray}
\int_0^\infty \frac{d C_t}{d \check r} k \check r^2 d \check r
&=& - \eta A \xi X (1 + \alpha_{eb}) r^{-\eta + 2}\nonumber \\
&& \left(  \frac{1}{-\eta + 2 + b (r)}  +  \frac{1}{\eta - 2} \right). \label{Eq:lim_Integral}
\end{eqnarray}

Here, we also assume a constant slope of the equilibrium state. To give this assumption, we find $b(r)$ such that
\begin{eqnarray}
\alpha_{sc} r^\beta = \frac{1}{-\eta + 2 + b (r)}  +  \frac{1}{\eta - 2}, \label{Eq:alphasc}
\end{eqnarray}
where $\alpha_{sc}$ and $\beta$ are constant. This form yields 
\begin{eqnarray}
b(r) = \frac{ a_{sc} r^{\beta} (\eta - 2)^2}{\alpha_{sc} r^{\beta} (\eta - 2) -1}. \label{Eq:b(r)}
\end{eqnarray}
Using this $\beta$ value, we rewrite Equation (\ref{Eq:lim_Integral}) as
\begin{eqnarray}
\int_0^\infty \frac{d C_t}{d \check r} k \check r^2 d \check r
&=& - \eta A \xi X (1 + \alpha_{eb}) \alpha_{sc} r^{-\eta + 2 + \beta}. \label{Eq:lim_Integral2}
\end{eqnarray}

\begin{figure}[!]
  \centering
  \includegraphics[width=3in]{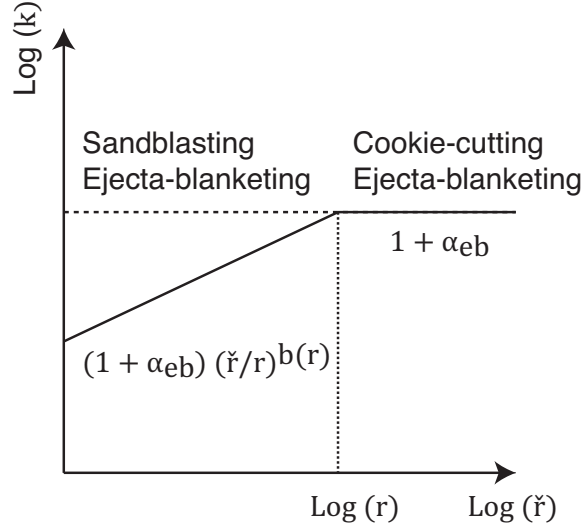}
  \caption{Schematic plot of the degradation parameter in log-log space. The $x$ axis indicates $\check r$ in a log scale, while the $y$ axis shows the value of the degradation parameter in a log scale. If $\check r \ge r$, $k$ is always $1 + \alpha_{eb}$ because cookie-cutting and ejecta-blanketing are dominant. If $\check r < r$, sandblasting and ejecta-blanketing are considered to be dominant. For this case, $k$ is described as $(1 + \alpha_{eb}) (\check r/r)^{b(r)}$. The slope, $b(r)$, changes as a function of $r$.}
  \label{Fig:AlphaBeta}
\end{figure}

\subsection{Equilibrium state}
This section introduces the equilibrium slope at $r_{min} \rightarrow 0$ and $r_{max} \rightarrow \infty$. Impact cratering achieves its equilibrium states on a surface when $t \rightarrow \infty$. Using Equations (\ref{Eq:C_c}) and (\ref{Eq:lim_Integral2}), we write an ordinal differential equation of the equilibrium state as
\begin{eqnarray}
\frac{d C_{c}^\infty}{d r} &=& - \frac{\frac{d \dot C_t}{d r}}{\frac{\pi}{A q} \int_{0}^{\infty} \frac{d \dot C_t}{d \check r} k \check r^2 d \check r}, \nonumber \\
&=& - \frac{A q}{\pi (1 + \alpha_{eb}) \alpha_{sc}} r^{-3 - \beta}. \label{Eq:Equilibrium}
\end{eqnarray}
Integrating Equation (\ref{Eq:Equilibrium}) from $r$ to $\infty$, we derive the visible crater CSFD at the equilibrium condition as
\begin{eqnarray}
C_{c}^\infty &=& - \int_{r}^\infty \frac{d C_c}{d r} dr, \nonumber \\
&=& \frac{A q}{\pi (1 + \alpha_{eb}) \alpha_{sc} (2 + \beta)} r^{- 2 - \beta}. \label{Eq:EqC_c}
\end{eqnarray}

Equation (\ref{Eq:EqC_c}) indicates that the fraction term on the right-hand side is independent of $\xi$ and $\eta$, and the equilibrium slope is simply given as $2 + \beta$. If $\beta = 0$, the equilibrium slope is exactly 2. This statement results from a constant value of $b(r)$, the case of which was discussed by \cite{Marcus1970} and \cite{Soderblom1970}. These results indicate that the equilibrium state is independent of the crater production function and only dependent on the surface condition. Therefore, a better understanding of the degradation parameter may provide strong constraints on the properties of cratered surfaces, such as regional slope effects, material conditions, and densities.  

We briefly explain the case of $\eta - 2 < 0$. This would be a ``shallow-sloped'' CSFD, such as seen in large crater populations on heavily-cratered ancient surfaces. For this case, cookie-cutting is the primary process that erases old craters \citep{Richardson2009}. Considering that $r_{min} \rightarrow 0$ and $r_{max}$ becomes quite large ($\gg r)$, we use Equation (\ref{Eq:integral}) to approximately obtain
\begin{eqnarray}
\int_{0}^{r_{max} \gg r} \frac{d C_t}{d \check r} k \check r^2 d \check r \propto r_{max}^{-\eta + 2},
\end{eqnarray} 
which is constant. Thus, from Equation (\ref{Eq:EqC_c}), we derive
\begin{eqnarray}
C_{c}^\infty \propto C_t. 
\end{eqnarray}
This equation means that the slope of the equilibrium state is proportional to that of the produced crater CSFD, which is consistent with the arguments by \cite{Chapman1986} and \cite{Richardson2009}. We leave detailed modeling of this case as a future work.

\subsection{Time evolution of countable craters}
This section calculates the time evolution of the visible crater CSFD, $C_c$. Substituting Equation (\ref{Eq:EqC_c}) into Equation (\ref{Eq:C_c}) yields 
\begin{eqnarray}
\frac{d C_c}{d r} &=& - \frac{A q}{\pi (1 + \alpha_{eb}) \alpha_{sc}} r^{- 3 - \beta} \left[ 1 - \exp \left \{ - \frac{\pi \eta \xi X}{q} (1 + \alpha_{eb}) \alpha_{sc} r^{-\eta + 2 + \beta} \right \} \right]. \label{Eq:partialCc_final}
\end{eqnarray}
Integrating Equation (\ref{Eq:partialCc_final}) from $r$ to $\infty$, we obtain 
\begin{eqnarray}
C_{c} &=& - \int_r^\infty \frac{d C_c}{d r} dr, \nonumber \\
&=& \frac{A q}{\pi (1 + \alpha_{eb}) \alpha_{sc} (2 + \beta)} r^{- 2 - \beta} \label{Eq:C_c_incomp} \\
&+& \frac{A q}{\pi (1 + \alpha_{eb}) \alpha_{sc}} \int^\infty_r r^{- 3 - \beta} \exp \left \{ - \frac{\pi \eta \xi X}{q} (1 + \alpha_{eb}) \alpha_{sc} r^{-\eta + 2 + \beta} \right \} dr.  \nonumber
\end{eqnarray}

While the second row in this equation directly results from Equation (\ref{Eq:EqC_c}), the third row needs additional operations. To derive the analytical form of the integral term, we introduce an incomplete form of the gamma function, which is given as
\begin{eqnarray}
\Gamma (a, Z) = \int_Z^\infty Z^{a - 1} \exp (-Z) dZ. \label{Eq:Gamma}
\end{eqnarray}
We focus on the critical integral part of Equation (\ref{Eq:C_c_incomp}), which is given as 
\begin{eqnarray}
f &=& \int_r^\infty r^{- 3 - \beta} \exp \left( - \chi r^{-\eta + 2 + \beta} \right) dr, \label{Eq:f1}
\end{eqnarray}
where
\begin{eqnarray}
\chi &=& \frac{\pi \eta \xi X}{q} (1 + \alpha_{eb}) \alpha_{sc}. \label{Eq:chi}
\end{eqnarray}

To apply Equation (\ref{Eq:Gamma}) to Equation (\ref{Eq:f1}), we consider the following relationships,
\begin{eqnarray}
Z &=& \frac{\chi}{r^{\eta - 2 - \beta}}, \label{Eq:t} \\
r &=& \left( \frac{\chi}{Z} \right)^\frac{1}{\eta-2-\beta}. \label{Eq:rForZ}
\end{eqnarray}
Equation (\ref{Eq:rForZ}) provides
\begin{eqnarray}
dr = \frac{-1}{\eta - 2 - \beta} \left( \frac{\chi}{Z} \right)^{\frac{1}{\eta - 2 - \beta}} \frac{dZ}{Z}. \label{Eq:dr}
\end{eqnarray}
Using Equations (\ref{Eq:t}) through (\ref{Eq:dr}), we describe $f$ as
\begin{eqnarray}
f &=& - \frac{1}{\eta - 2 - \beta} \left( \frac{1}{\chi} \right)^{\frac{2 + \beta}{\eta - 2 - \beta}} \int_Z^0 Z^{\frac{2 + \beta}{\eta - 2 - \beta} - 1}\exp (- Z) dZ,  \nonumber \\
&=& 
\frac{1}{\eta - 2 - \beta} \left( \frac{1}{\chi} \right)^{\frac{2 + \beta}{\eta - 2 - \beta}} \gamma \left( \frac{2 + \beta}{\eta - 2 - \beta}, Z \right), 
\end{eqnarray}
where $\gamma(\cdot, 0)$ is called a lower incomplete gamma function. For the current case, this function is defined as
\begin{eqnarray}
\gamma \left( \frac{2 + \beta}{\eta - 2 - \beta}, Z \right) = \Gamma \left( \frac{2  + \beta}{\eta - 2 - \beta} \right) -  \Gamma \left( \frac{2 + \beta}{\eta - 2 - \beta}, Z \right),
\end{eqnarray}
where $\Gamma (\cdot) = \Gamma (\cdot, 0)$. We eventually obtain the final solution as
\begin{eqnarray}
C_c &=&  \frac{A q}{\pi (1 + \alpha_{eb}) \alpha_{sc}} \frac{1}{2 + \beta} r^{- 2 - \beta} \label{Dsd} \\
&+&  \frac{A q}{\pi (1 + \alpha_{eb}) \alpha_{sc}}  \frac{1}{\eta - 2 - \beta} \left( \frac{1}{\chi} \right)^{\frac{2 + \beta}{\eta - 2 - \beta}} \gamma \left( \frac{2 + \beta}{\eta - 2 - \beta}, Z \right) . \nonumber
\end{eqnarray}
At an early stage, both the first term and the second term play a role in determining $C_c$, which should be close to the produced crater CSFD. However, as the time increases, $X$ also becomes large. From Equation (\ref{Eq:f1}), when $X \gg 1$, $f \ll 1$, and thus the second term becomes negligible. This process causes $C_c$ to become close to $C^\infty_c$.

We introduce a special case of $\beta = 0$ and $\eta = 3$. From Equation (\ref{Eq:b(r)}), $b(r)$ becomes constant and is given as
\begin{eqnarray}
b = \frac{\alpha_{sc}}{\alpha_{sc} - 1}. \label{Eq:bsp}
\end{eqnarray}
Equation (\ref{Dsd}) is simplified as 
\begin{eqnarray}
C_c &=&  \frac{A q}{2 \pi (1 + \alpha_{eb}) \alpha_{sc}} r^{- 2} +  \frac{A q}{\pi (1 + \alpha_{eb}) \alpha_{sc}} \chi^{-2} \gamma (2 , Z), \nonumber \\
&=& \frac{A q}{2 \pi (1 + \alpha_{eb}) \alpha_{sc}} r^{- 2} +  \frac{A q}{\pi (1 + \alpha_{eb}) \alpha_{sc}} \chi^{-2} \left \{ 1 - \left( 1 + \frac{\chi}{r} \right) \exp \left( - \frac{\chi}{r} \right) \right \}. 
\end{eqnarray}
Equation (\ref{Eq:bsp}) shows that when $\alpha_{sc} \rightarrow 1$, $b \rightarrow \infty$. However, the following exercises show that $\alpha_{sc} \gg 1$. 

\section{Sample applications}
\label{Sec:Behavior}
We apply the developed model to the visible crater CSFD of the Sinus Medii region on the Moon (see the red-edged circles in Figure \ref{Fig:SinusMedii}) and that of the Apollo 15 landing site (see the red-edged circles in Figure \ref{Fig:AP15}). These locations are considered to have reached crater count equilibrium. In these exercises, we obtain $C_t$ by considering the crater sizes that have not researched equilibrium, yet. Then, assuming that $\alpha_{eb} = 0$, we determine $\alpha_{sc}$ such that $C_c$ matches the empirical datasets.

\subsection{The Sinus Medii region}
\cite{Gault1970} obtained this CSFD (see Figure 14 in his paper), using the so-called nesting counting method. This method accounts for large craters in a global region, usually obtained from low-resolution images, and small craters in a small region, given from high-resolution images. In the following discussion, caution must be taken to deal with the units of the degradation constants.

\begin{figure}[!]
  \centering
  \includegraphics[width=4in]{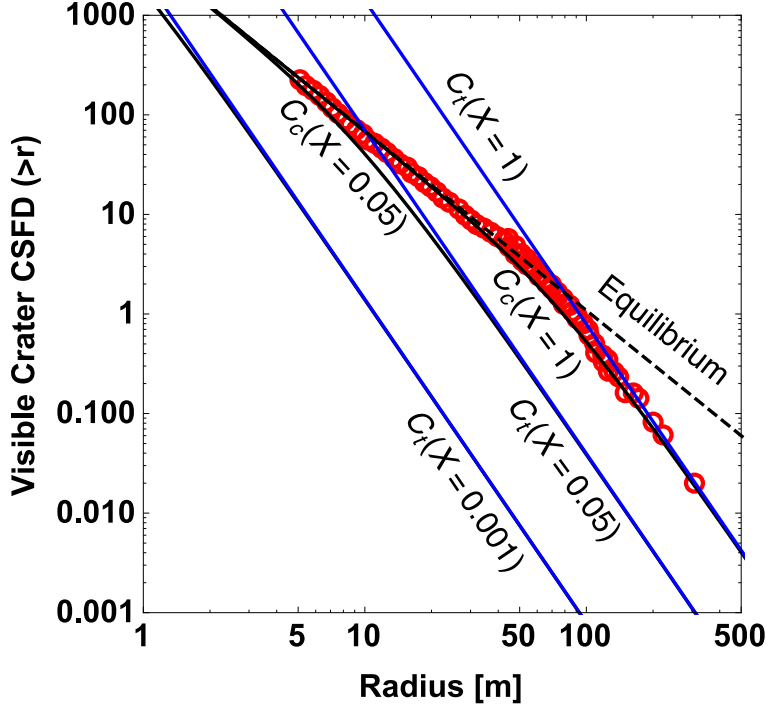}
  \caption{Comparison of the analytical results with the empirical data of the Sinus Medii region by \cite{Gault1970}. The area plotted is 1 km$^2$. The red-edged circles are the empirical data. The blue and black lines show the CSFDs of the produced craters, $C_t$, and that of the visible craters, $C_c$, respectively. We plot the results at three different times: $X = 0.001$, 0.05, and 1.0. The dashed line indicates the equilibrium condition.}
  \label{Fig:SinusMedii}
\end{figure}

Studies of the crater production function \cite[e.g.][]{Neukum2001} showed that a high slope region, which usually appears at sizes $\sim 100$ m to $\sim$1 km might be similar to the produced crater CSFD. Here, we observed that such a steep slope appears between $\sim$100 m and $\sim$400 m on the Sinus Medii surface. By fitting this high slope, we obtain $C_t$ as
\begin{eqnarray}
C_t  = 2.5 \times 10^6 r^{-3.25}. \label{Eq:C_t_SinusMedii}
\end{eqnarray}
$C_t$ for the Sinus Medii case is the produced crater CSFD for an area of 1 km$^2$ (we consider an area of 1 m$^2$ to be the unit area), and this quantity is dimensionless, and the $2.5 \times 10^6$ factor has units of m$^{3.25}$. Since $\eta = 3.25  > 2$, this case is the high-slope crater production function. Based on this fitting function, we set $X = 1$, $\xi = 2.5$ m$^{1.25}$, and $A = 1$ km$^2$. Also, we obtain the fitting function of the equilibrium slope as 
\begin{eqnarray}
C_c^\infty = 4.3 \times 10^3 r^{-1.8}. \label{Eq:C_c_SinusMedii}
\end{eqnarray}
Similar to $C_t$, the units of the $4.3 \times 10^3$ factor are m$^{1.8}$. Figure (\ref{Fig:SinusMedii}) compares the empirical data with the time evolution of $C_c$ that is given by Equation (\ref{Dsd}). We describe different time points by varying $X$ without changing $\xi$ and $A$. It is found that the model captures the equilibrium evolution properly. 

\begin{figure}[!]
  \centering
  \includegraphics[width=4in]{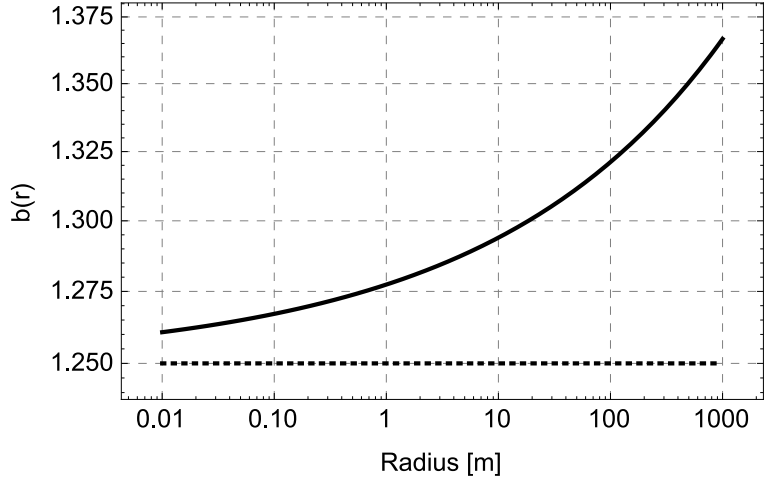}
  \caption{Variation in $b(r)$ for the Sinus Medii case. The solid line shows $b(r)$, which is given in Equation (\ref{Eq:b(r)}). The dotted line represents the minimum value of $b(r)$, which is $\eta - 2 = 1.25$ for the Sinus Medii case.  }
  \label{Fig:b(r)}
\end{figure}

By fitting $C_c$ with the empirical data, the present model can provide constraints on the sandblasting exponent, $b(r)$, of the degradation parameter. To obtain this quantity, we determine $\alpha_{sc}$ and $\beta$. Since $\alpha_{eb}$ is assumed to be negligible, we write $1 + \alpha_{eb} \sim 1$. First, since $- 2 - \beta = - 1.8$, we derive $\beta = - 0.2$. Second, operating the units of the given parameters, we have the following relationship,
\begin{eqnarray}
4.3 \times 10^3 \: \text{[m$^{1.8}$]} = \frac{A q}{\pi (2 + \beta) \alpha_{sc}} = \frac{10^6\: \text{[m$^2$]} \: \times 0.907}{\pi (2 - 0.2) \alpha_{sc} \: \text{[m$^{0.2}$]}}.
\end{eqnarray}
Then, we obtain 
\begin{eqnarray}
\alpha_{sc} = 37.3 \: \text{[m$^{0.2}$]}. 
\end{eqnarray}
Since the units of $r^{\beta}$ are m$^{-0.2}$, this result guarantees that Equation (\ref{Eq:b(r)}) consistently provides a dimensionless value of $b(r)$. Using these quantities, we obtain the variation in $b(r)$. Figure \ref{Fig:b(r)} indicates that the obtained values of $b(r)$ for the Sinus Medii satisfy the sand-blasting condition, $b(r) > \eta - 2$. For this case, which has a constant slope index, $b(r)$ monotonically increases. As newly emplaced craters become smaller, they become less capable of erasing a crater. These results imply that for a large simple crater, it would take longer time for smaller craters to degrade its deep excavation depth and its high crater rim. These results could be used to constrain models for net downslope material displacement by craters; however, this is beyond our scope in this paper.

\subsection{The Apollo 15 landing site}
\begin{figure}[!]
  \centering
  \includegraphics[width=4in]{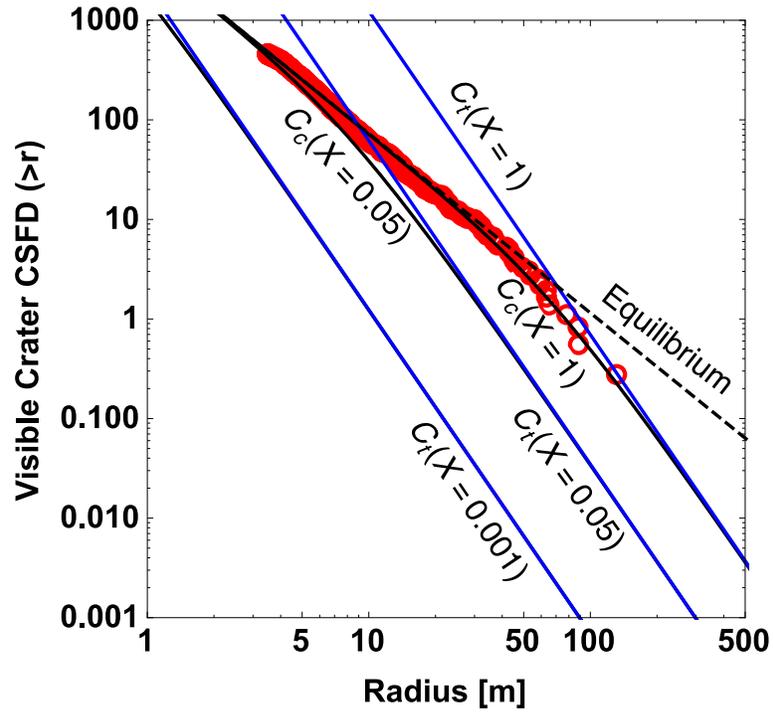}
  \caption{Comparison of the analytical results with the empirical data of the Apollo 15 landing region over an area of 1 km$^2$. C.I.F. counted craters on this region in \cite{Robbins2014}. The red-edged circles are the empirical data. The definitions of the line formats are the same as those in Figure \ref{Fig:SinusMedii}.}
  \label{Fig:AP15}
\end{figure}

We also consider the crater equilibrium state on the Apollo 15 landing site. Co-author Fassett counted craters at this area in \cite{Robbins2014}, and we directly use this empirical result. The used image is a sub-region of M146959973L taken by Lunar Reconnaissance Orbiter Camera Narrow-Angle Camera, the image size is 4107 $\times$ 2218 pixels, and the solar incidence angle is 77$^\circ$ \citep{Robbins2014}. The pixel size of the used image is 0.63 m/pixel. The total domain of the counted region is 3.62 km$^2$, and the number of visible craters is 1859. We plot the empirical data in Figure \ref{Fig:AP15}. To make this figure consistent with Figure \ref{Fig:SinusMedii}, we plot the visible crater CSFD with an area of 1 km$^2$. In the following discussion, we will show $C_t$, $C_c^\infty$, and $C_c$ by keeping this area, i.e., $A = 1$ km$^2$, to make comparisons of our exercises clear. 

\begin{figure}[!]
  \centering
  \includegraphics[width=4in]{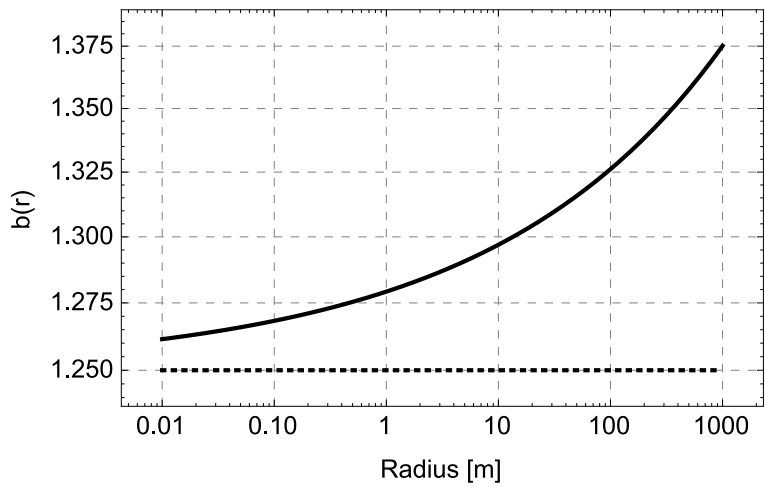}
  \caption{Variation in $b(r)$ for the Apollo 15 landing site case.}
  \label{Fig:b(r)AP15}
\end{figure}

For the Apollo 15 landing site, the steep-slope region ranges from $\sim 50$ m to the maximum crater radius, which is 131 m. The fitting process yields 
\begin{eqnarray}
C_t  = 2.2 \times 10^6 r^{-3.25}. \label{Eq:C_t_SinusMedii}
\end{eqnarray}
This fitting process shows that $C_t$ of the Apollo 15 landing site is consistent with that of the Sinus Medii case. Then, given $A = 1$ km$^2$, we obtain $\xi = 2.2$ m$^{1.25}$. We also set $X = 1$ for the condition that fits the empirical dataset. $C_c^\infty$ is given as
\begin{eqnarray}
C_c^\infty = 4.6 \times 10^3 r^{-1.8}. \label{Eq:C_c_SinusMedii}
\end{eqnarray}
The units of the $4.6 \times 10^3$ factor are m$^{1.8}$. Figure (\ref{Fig:AP15}) shows comparisons of the analytical model and the empirical data for the Apollo 15 landing site. We also obtain $\alpha_{sc}$ and $\beta$. Again, $\alpha_{eb}$ is assumed to be zero. Since the slope of the equilibrium state is 1.8, we derive $\beta = - 0.2$. Similar to the Sinus Medii case, we calculate $\alpha_{sc}$ for the Apollo 15 case as 34.9 m$^{-0.2}$. These quantities yield the variation in $b(r)$ (Figure \ref{Fig:b(r)AP15}). The results are consistent with those for the Sinus Medii case.

\section{Necessary improvements}
Further investigations and improvements will be necessary as we made five assumptions in the present study. First, we simply considered the limit condition of the crater radius, i.e., $r_{min} \rightarrow 0$ and $r_{max} \rightarrow \infty$. However, this assumption neglects consideration of a cut-off effect on crater counting. Such an effect may happen when a local area is chosen to count craters on a terrain that reaches crater count equilibrium. Due to this effect, large craters in the area may be accidentally truncated, and the craters counted there may not follow the typical slope feature. Second, we ignored the effect of ejecta-blanketing in our exercises. However, it is necessary to investigate the details for it to give stronger constraints on the crater count equilibrium problem. Third, we assumed that the equilibrium state is characterized by a single slope. However, an earlier study has shown that the equilibrium slope could vary at different crater sizes from case to case \cite[e.g.][]{Robbins2014}. To adapt such complex equilibrium slopes, we require more sophisticated forms of Equation (\ref{Eq:alphasc}). Fourth, the current version of this model does not distinguish crater degradation with crater obliteration. For example, if $t \ll 1$ in Equation (\ref{Dsd}), there is a chance that craters would be degraded due to sandblasting but not obliterated. At this condition, they all would be visible, while Equation (\ref{Dsd}) predicts some obliteration. Fifth, the measured crater radius can increase due to sandblasting, while the current model does not account for this effect. We will attempt to solve these problems in our future works. 

We finally address that although we took into account cookie-cutting, ejecta-blanketing, and sandblasting as the physical processes contributing to crater count equilibrium in this study, we have not implemented the effect of crater counting on the degradation parameter. According to \cite{Robbins2014}, the visibility of degraded craters could depend on several different factors: sharpness of craters, surface conditions, and image qualities (such as image resolution and Sun angles). Also, purposes that a crater counter has also play a significant role in crater counting. A better understanding of this mechanism will shed light on the effect of human crater counting processes on crater count equilibrium. We will conduct detailed investigations and construct a better methodology for characterizing this effect.

\section{Conclusion}
We developed an analytical model for addressing the crater count equilibrium problem. We formulated a balance condition between crater accumulation and crater degradation and derived the analytical solution that described how the crater count equilibrium evolves over time. The degradation process was modeled by using the degradation parameter that gave an efficiency for a new crater to erase old craters. This model formulated cookie-cutting, ejecta-blanketing, and sandblasting to model crater count equilibrium. 

To formulate the degradation parameter, we considered the slope functions of the ratio of one crater to the other for the following cases: if the size of newly emplaced craters was smaller than that of old craters, ejecta-blanketing and sandblasting were dominant; otherwise, ejecta-blanketing and cookie-cutting mainly erased old craters. Based on our formulation of the degradation parameter, we derived the relationship between this parameter and a fitting function obtained by the empirical data. If the slope of the crater production function was higher than 2, the equilibrium state was independent of the crater production function. We recovered the results by earlier studies that the slope of the equilibrium state was always independent of the produced crater CSFD. If the physical processes were scale-dependent, the slope deviated from the slope of 2. 

Using the empirical results of the Sinus Medii region and the Apollo 15 landing site on the Moon, we discussed how our model constrained the degradation parameters from observed crater counts of equilibrium surfaces. We assumed that the ejecta-blanketing process was negligible. This exercise showed that this model properly described the nature of crater count equilibrium. Further work will be conducted to better understand the slope functions of the degradation parameters, which will help us do validation and verification processes for both our analytical model and the numerical cratered terrain model CTEM \citep{Richardson2009, Minton2015}. 

\section{Acknowledgements}
M.H. is supported by NASA's GRAIL mission and NASA Solar System Workings \\ NNX15AL41G. The authors acknowledge Dr. Kreslavsky and the anonymous reviewer for detailed and useful comments that substantially improved our manuscript. The authors also thank Dr. H. Jay Melosh at Purdue University, Dr. Jason M. Soderblom at MIT, Ms. Ya-Huei Huang at Purdue University, and Dr. Colleen Milbury at West Virginia Wesleyan College for useful advice to this project.

%\bibliographystyle{elsarticle-harv}
%\bibliography{eqm_crater} 

\end{document}